\documentclass[twocolumn]{emulateapj}
\usepackage{longtable}
\usepackage{graphics}
\usepackage{rotating}
\usepackage{amsmath}
\usepackage{txfonts}
\usepackage{hyperref}
\usepackage{color}
\def\gtorder{\mathrel{\raise.3ex\hbox{$>$}\mkern-14mu
             \lower0.6ex\hbox{$\sim$}}}
\def\ltorder{\mathrel{\raise.3ex\hbox{$<$}\mkern-14mu
             \lower0.6ex\hbox{$\sim$}}}
\slugcomment{\today}
\shorttitle{GW170817 radio polarization}
\shortauthors{Corsi et al.}

\begin{document}
\title{An upper-limit on the linear polarization fraction of the GW170817 radio continuum}
\author{Alessandra~Corsi\altaffilmark{1}}
 \author{Gregg~W.~Hallinan\altaffilmark{2}}
  \author{Davide~Lazzati\altaffilmark{3}}
  \author{Kunal~P.~Mooley\altaffilmark{4,2,5}}
 \author{Eric~J.~Murphy\altaffilmark{6}}
 \author{Dale~A.~Frail\altaffilmark{4}}
 \author{Dario~Carbone\altaffilmark{1}}
 \author{David~L.~Kaplan\altaffilmark{7}}
 \author{Tara~Murphy\altaffilmark{8}}
 \author{Shrinivas~R.~Kulkarni\altaffilmark{2}}
 \author{Kenta~Hotokezaka\altaffilmark{9}}
 \altaffiltext{1}{Department of Physics and Astronomy, Texas Tech University, Box 1051, Lubbock, TX 79409-1051, USA; e-mail: alessandra.corsi@ttu.edu}
 \altaffiltext{2}{Caltech, 1200 E. California Blvd. MC 249-17, Pasadena, CA 91125, USA}
 \altaffiltext{3}{Department  of  Physics,  Oregon  State  University, 301  Weniger  Hall,  Corvallis,  OR,  97331,  USA}
 \altaffiltext{4}{National Radio Astronomy Observatory (NRAO), P.O. Box O, Socorro, New Mexico 87801, USA}
 \altaffiltext{5}{Jansky Fellow (NRAO, Caltech)}
 \altaffiltext{6}{National Radio Astronomy Observatory
520 Edgemont Road Charlottesville, VA 22903, USA}
\altaffiltext{7}{Department of Physics, University of Wisconsin - Milwaukee, Milwaukee, Wisconsin 53201, USA}
 \altaffiltext{8}{Sydney Institute for Astronomy, School of Physics, University of Sydney, Sydney, New South Wales 2006, Australia}
 \altaffiltext{9}{Department Astrophysical Sciences, Princeton University, Peyton Hall, Princeton, NJ 08544, USA}
 
\begin{abstract}
We present late-time radio observations of GW170817, the first binary neutron star merger discovered through gravitational waves by the advanced LIGO and Virgo detectors. Our observations, carried out with the Karl G. Jansky Very Large Array, were optimized to detect polarized radio emission, and thus to constrain the linear polarization fraction of GW170817. At an epoch of $\approx 244$\,days after the merger, we rule out linearly polarized emission above a fraction of $\approx 12$\% at a frequency of 2.8\,GHz  (99\% confidence). Within the structured jet scenario (a.k.a. successful jet plus cocoon system) for GW170817, the derived upper-limit on the radio continuum linear polarization fraction strongly constrains the magnetic field configuration in the shocked ejecta. We show that our results for GW170817 are compatible with the low level of linear polarization found in afterglows of cosmological long gamma-ray bursts. Finally, we discuss our findings in the context of future expectations for the study of radio counterparts of binary neutron star mergers identified by ground-based gravitational-wave detectors.
\end{abstract}

\section{Introduction}
\label{sec:intro}
On August 17$^{\rm th}$ 2017, the field of gravitational-wave (GW) astronomy reached the big leagues with a dazzling discovery. Only $\approx 8$\,d before the official end of their second observing run (O2), advanced LIGO and Virgo obtained their first direct detection of GWs from a binary neutron-star (NS) merger, an event dubbed GW170817 \citep{Abbott2017a}.  

After the GW discovery, GW170817 gifted the astronomical community with an electromagnetic (EM) counterpart spanning all bands of the spectrum \citep[e.g.,][and references therein]{Abbott2017b}. 
Only $\approx 2$\,s after the GW170817 merger, a short $\gamma$-ray burst (GRB) was detected by the \textit{Fermi} and \textit{Integral} satellites \citep{Abbott2017c,Goldstein2017,Savchenko2017}. The discovery of $\gamma$-rays from GW170817 was followed by the detection of a UV/optical/IR counterpart in NGC~4993, a lenticular galaxy located at $\approx 40$\,Mpc \citep{Coulter2017}, resulting in the nearest short GRB with a measured redshift \citep[e.g.,][]{Fong2017}. While it had long been thought that short GRBs are NS-NS mergers launching relativistic jets pointed directly at us, the close distance of GW170817 implied that its $\gamma$-ray counterpart was $\sim10^3-10^4\times$ less energetic than the previously known population of short GRBs \citep{Fong2017,Goldstein2017}. This formed the first piece of an intriguing puzzle.

The early UV/optical/IR emission from GW170817 was rather different from the non-thermal optical afterglows of short GRBs, and soon recognized to be dominated by a ``kilonova'', a quasi-thermal transient powered by the radioactive decay of r-process nuclei \citep[e.g.,][]{Arcavi2017,Chornock2017,Cowperthwaite2017,Drout2017,Evans2017,Kasliwal2017,Kasen2017,Kilpatrick2017,Pian2017,Shappee2017,Smartt2017,Tanvir2017,Valenti2017}. This kilonova detection solved a decades-old mystery of where most of the elements heavier than iron are synthesized \citep[for a recent review, see e.g.,][and references therein]{Metzger2017}.

About 10 days after the GW discovery of GW170817, an X-ray counterpart was detected by the \textit{Chandra} satellite \citep{Haggard2017,Margutti2017,Troja2017}. A delayed radio afterglow was unveiled by the Karl G. Jansky Very Large Array (VLA) about two weeks after the merger \citep{Hallinan2017}, and subsequently confirmed by the Australia Telescope Compact Array (ATCA). Radio and X-ray observations of GW170817 probe a completely different emission mechanism than the kilonova observed at optical-IR wavelengths, namely, non-thermal radiation from the fastest ejecta. 
Continued VLA and ATCA monitoring over the first $\sim 100$\,d since the merger revealed a steady increase of the optically-thin synchrotron radio emission, followed by a turnover \citep[$t\gtrsim 150$\,d since merger;][]{Alexander2017,Alexander2018,Margutti2018,Mooley2018,Dobie2018}. Together with the weak $\gamma$-rays, these observations clearly set GW170817 apart from the previously known population of short GRBs with fast-fading afterglows \citep{Fong2017}. 
The relatively slow temporal rise of the radio flux \citep{Hallinan2017,Mooley2018,Dobie2018}, in particular, has ruled out the simplest scenario relating GW170817 radio counterpart to synchrotron emission from a uniform (``top-hat'') jet shocking the ISM \citep[][]{Margutti2017,Troja2017}, as usually invoked to explain cosmological GRB afterglows \citep[e.g.,][]{Sari1999}.

\begin{table*}
\caption{Sensitivity reached in our VLA polarization observations of GW170817. \label{Tab:1}}
\begin{center}
\begin{tabular}{cccccc}
\hline\hline
UTC & Epoch & $\Delta T_{\rm obs}$ & $\nu$  & Stokes Q rms & Stokes U rms \\
    & (days since 2017 Aug 17.528 UTC) &  (hr)        & (GHz)        & ($\mu$Jy/beam) & ($\mu$Jy/beam)  \\
    \hline
  2018 Mar 02.321  & 197 &  1.5 & 3.0 & 4.5&  4.5 \\
  2018 Mar 25.344  
  & 220 & 3.5 & 2.8  & 3.3 & 3.3 \\
  2018 Mar 26.310 
  & 221& 3.5 & 2.8 & 3.4 & 3.4 \\
  2018 May 11.167  
  & 267& 3.5 & 2.8 & 3.5 & 3.4\\
  2018 May 12.168  
  & 268 & 3.5 & 2.8 & 3.4 & 3.4 \\
  2018 Mar 25 - May 12 
  & $244\pm24$& $3.5\times4$ & 2.8 & 1.7& 1.7\\
  \hline
\end{tabular}
\end{center}
\end{table*}

Broadly speaking, two main scenarios have been proposed to explain GW170817 non-thermal emission: (i) A successful structured jet (a.k.a. successful jet - cocoon system) composed of an outflow with a narrow, highly relativistic core (similar to cosmological GRB jets) initially directed away from our line of sight (off-axis), plus slower-moving wings \citep[e.g.,][]{Lazzati2017a,Lazzati2017b,Lazzati2018,Margutti2018}; (ii) A choked-jet scenario where the jet is unable to break out of the neutron-rich dynamical ejecta, and the bulk of the energy is imparted to a radially stratified and quasi-spherical (or wide-angle) mildly relativistic cocoon \citep{Kasliwal2017,Gottlieb2018,Nakar2017,Mooley2018,Nakar2018a}. We note that a dynamical ejecta model where radio emission arises from the fast tail of the dynamical merger ejecta has also been proposed \citep[e.g.,][]{Hotokezaka2018}, but is somewhat unlikely given the sharp peak and fast decline of the radio light curve \citep[e.g.,][]{Dobie2018,Alexander2018}. Thus, in what follows, we will not discuss this third scenario further. 

While scenarios (i) and (ii) above imply very different geometries for the outflow, there are sufficient free model parameters that it has not been possible to distinguish them based on the {\it total radio intensity} alone \citep[e.g.,][]{Margutti2018}. Fortunately, detecting a polarized radio signal from GW170817 can provide a useful discriminant. Indeed, the presence of a large degree $(\approx 20\%)$ of linear polarization in the radio continuum would be a ``smoking gun'' for a high degree of asymmetry, and hence favor the jet scenario (i) \citep[e.g.,][]{Rossi2004,Gill2018,Nakar2018b,Lazzati2018}. In the absence of substantial linearly polarized emission, even though scenario (i) cannot be ruled out, strong constraints can be set on the structure of the post-shock magnetic field \citep[e.g.,][]{Ghisellini1999,Sari1999,Rossi2004,Gill2018,Nakar2018b}.

Motivated by the above considerations, here we present polarization observations of GW170817 radio counterpart. Our upper limit rules out strong linearly polarized GHz emission and, within the structured jet scenario, sets stringent constraints on the structure of the magnetic field within the shocked ejecta. Our paper is organized as follows. In Section \ref{sec:data} we describe our observations and data reduction; in Section \ref{sec:results} we discuss our results; in Section \ref{sec:conclusion} we summarize and conclude.

\section{Radio observations and data reduction}
\label{sec:data}
We observed the field of GW170817 with the Karl G. Jansky Very Large Array (VLA) in its most extended (A) configuration on multiple epochs between 2018 March 02.321 UTC and 2018 May 12.168 UTC (via projects VLA/17B-397 - PI: Mooley; and VLA/18A-457 - PI: Corsi). These observations were carried out in S-band, at a nominal central frequency of $\approx 3$\,GHz, and with a nominal 2\,GHz bandwidth. We included bandpass, flux density, and polarization position angle calibration scans on 3C286. The unpolarized source J1407$+$2827 was observed to calibrate for polarization leakage. During all epochs, J1248$-$1959 was used as our phase calibrator. 

The VLA data were first calibrated using the VLA automated calibration pipeline available in CASA, which is designed for Stokes I continuum calibration. Because our phase calibrator J1248$-$1959 is marginally resolved at the longest baselines of the VLA in its A configuration, we restricted the UV-range for this calibrator to $\lesssim 200$\,k$\lambda$ in all gain calibrator calls within the automated calibration pipeline. After the automated calibration, we set the polarization model for our polarized calibrator 3C286 \citep[11.2\% fractional polarization, and polarization angle of 33\,deg;][]{Perley2013}. Then, polarization calibration steps were carried out using the automated pipeline calibration tables for pre-calibration.  Specifically, we first solved for the cross-hand (RL, LR) delays due to the residual delay difference on the reference antenna used for the original delay calibration.
Then, we solved for the instrumental polarization (the frequency-dependent leakage terms, also referred to as ``D-terms'') using the unpolarized source J1407$+$2827. For all our observations, we found leakages $\lesssim 15\%$ for most antennas and spectral windows. Having calibrated the instrumental polarization, we carried out a frequency-dependent position angle calibration using the source 3C286. We found that the residual R-L phase on the reference antenna (after taking out the cross-hand delays) spanned about $10-15$\,deg across most spectral windows. 

After calibrating and visually inspecting the data for any further flagging, we run the CLEAN algorithm \citep{Hogbom1974} in interactive mode to image the fully calibrated Stokes IQUV, and derive single-epoch sensitivities. In the cleaning process, we used a  natural weighting of the visibilities so as to maximize the map point source sensitivity. We applied the same source mask to all polarizations, and used a pixel of size 0.2\,\arcsec (so as to oversample the synthesized beam, which was of $\approx 0.85\,\arcsec\times1.5\,\arcsec$ in our observations). A summary of our results is reported in Table \ref{Tab:1}. For each observation, we give the central UTC, the epoch in days since GW170817 merger time (2017 Aug 17.528 UTC 
), the total duration of the observation (including calibration), the central observing frequency, and the rms sensitivity reached in Stokes Q, U (we do not discuss Stokes V here, i.e. circular polarization, since no emission is expected or seen in this polarization state).  In Table \ref{Tab:1} we also report the sensitivity reached by co-adding our last four observations (which have comparable rms), and imaging the resulting dataset following the same procedure described above for the single epochs. The rms sensitivity reached in Stokes Q and U after co-adding is $\approx 1.7\,\mu$Jy/beam.

Using a circular region centered on the position of GW170817 \citep[$\alpha=13^{\rm h}09^{\rm m}48^{\rm s}.071$ and $\delta=-23^{\circ}22'53.37''$, J2000; e.g., ][]{Hallinan2017,Kasliwal2017} and of area comparable to that of the FWHM synthesized beam, we calculate the peak brightness measured in Stokes Q and Stokes U at the various epochs, and in the co-added dataset. In all cases we find that the measured Stokes Q and U peak brigtness at the GW170817 location is below $<3\times\sigma_{\rm Q,U}$ where $\sigma_{\rm Q,U}$ is the map rms. Thus, all our polarization observations yielded non-detections in Stokes Q and U.

\begin{figure}
\begin{center}
\vbox{
\includegraphics[width=9.1cm]{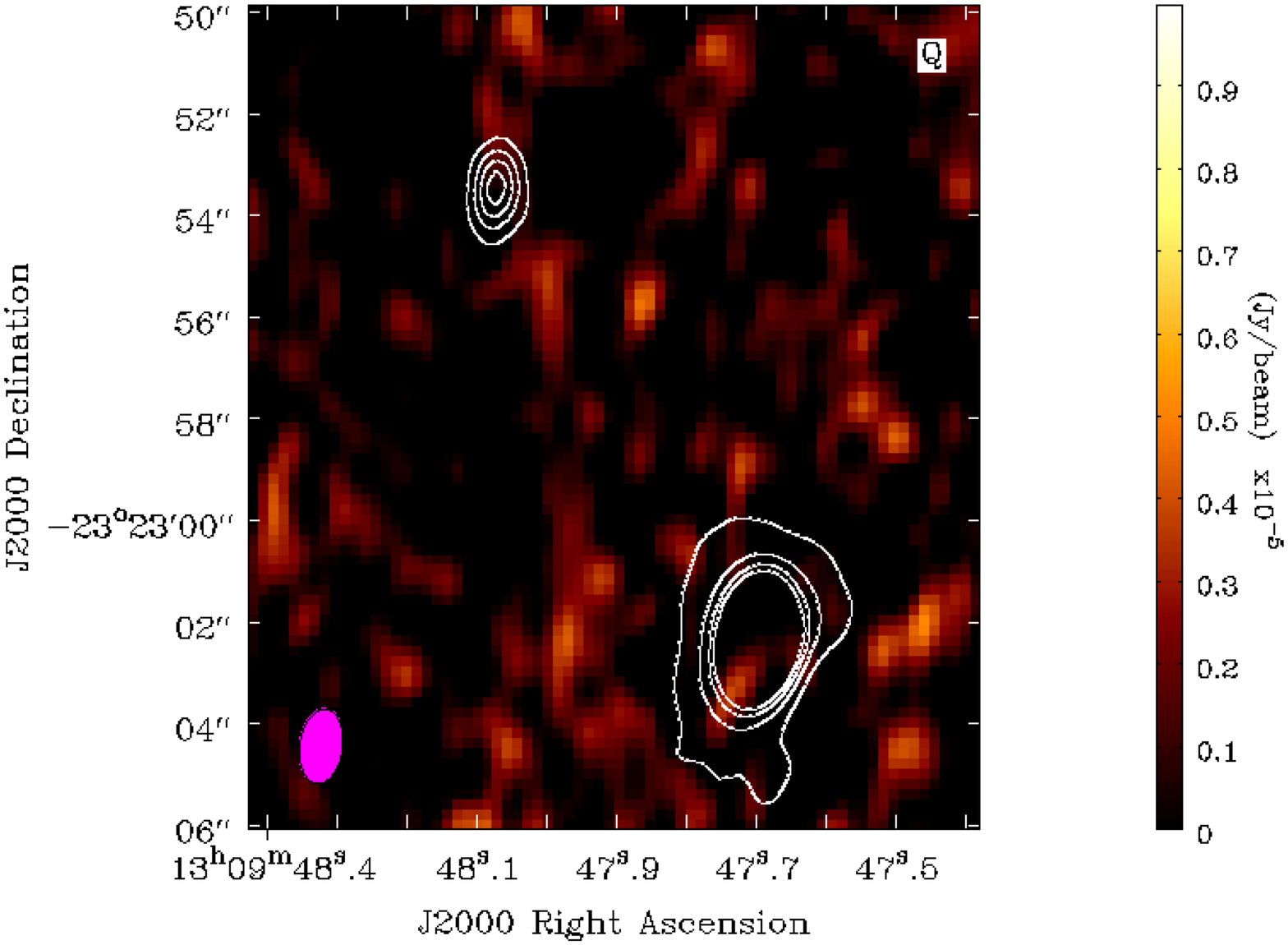}
\includegraphics[width=9.1cm]{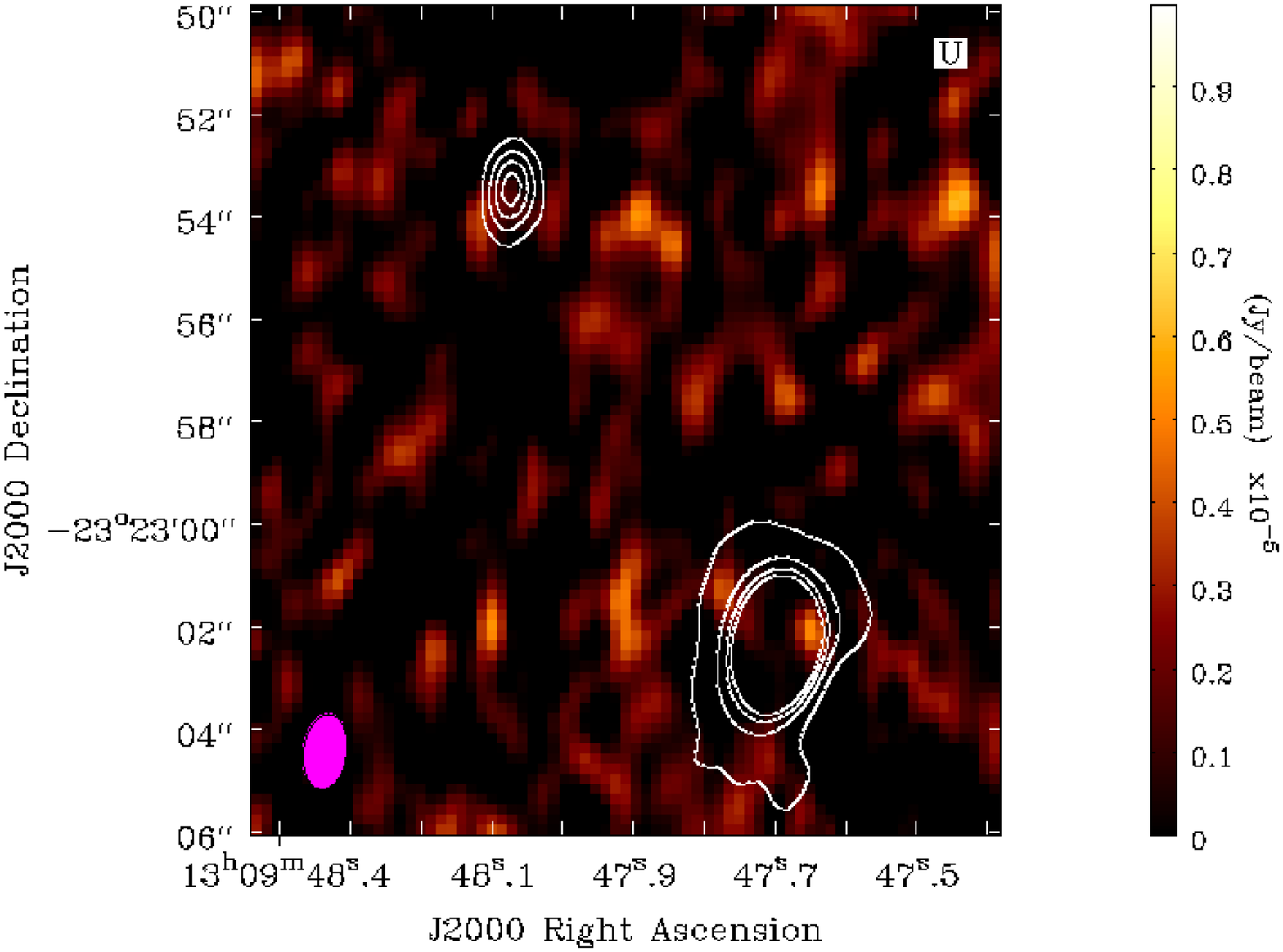}
}
\caption{TOP: Stokes Q intensity map of the co-added observations of the GW170817 field carried out in S-band between March 25 and May 12 (see Table 1). Stokes I contours of GW170817 radio counterpart are also shown (white; $20\%$, $40\%$, $60\%$, and $80\%$ relative emission contours). GW170817 radio counterpart is located at $\alpha=13^{\rm h}09^{\rm m}48^{\rm s}.071$, $\delta=-23^{\circ}22'53.37''$ \citep[J2000; e.g., ][]{Hallinan2017,Kasliwal2017}. The Stokes I intensity contours of the host galaxy of GW170817 are also overlaid (bottom-right portion of the panel). The FWHM synthesized beam ellipse is shown in magenta. BOTTOM: Same as the top panel, but for the Stokes U intensity map. \label{Fig:map}}
\end{center}
\end{figure}

\section{Results and discussion}
\label{sec:results}
\begin{figure}
\begin{center}
\hspace{-0.5cm}
\includegraphics[width=9.3cm]{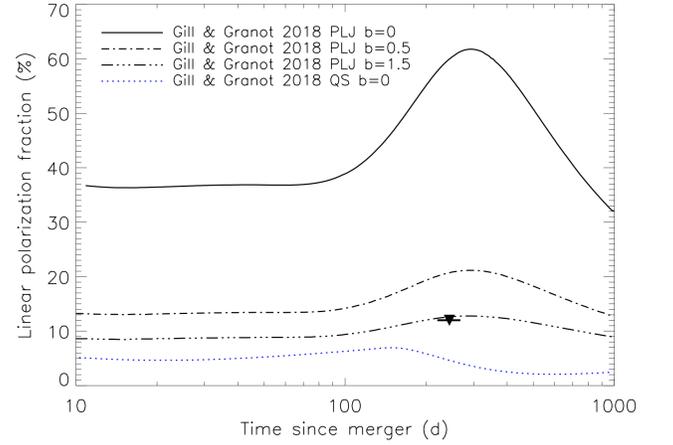}
\vspace{-0.5cm}
\caption{VLA upper-limit on the linear polarization fraction
$\sqrt{Q^2 + U^2}/I$ of the GHz radio flux of GW170817 (downward pointing triangle) compared with different theoretical predictions for the power-law structured jet model (PLJ; black), and for a quasi-spherical ejecta (QS; blue). These predictions are by \citet{Gill2018}. For the models here plotted, $b=0$ represents the case of a magnetic field completely contained in the plane of the shock, while $b>0$ is for a magnetic field whose component in the direction of the shock normal also
contributes. See text for discussion.\label{Fig:pol}}
\end{center}
\end{figure}

In our March 02 UTC observation, which had the shortest duration (Table 1), we measure $p=\sqrt{Q^2+U^2}/\sigma_{\rm U,V}\approx 3.0$ (where $\sigma_{U,V}=4.5\,\mu$Jy/beam; see Table 1) at 2.8\,GHz. Accounting for Ricean bias, we thus set a 99\% confidence upper limit of $p<5.2$ \citep{Vaillancourt2006}. The Stokes I peak brightness measured at this epoch is $(75.9\pm6.4)\,\mu$Jy (this includes a $5\%$ absolute flux density calibration error), and thus the corresponding upper-limit on the linear polarization fraction is $\Pi=\sqrt{Q^2+U^2}/I\lesssim 31\%$ at $\approx 197$\,d since merger. 

From the co-added map derived using our last four observations with comparable rms sensitivity (Table 1 and Fig. \ref{Fig:map}), we get $p=\sqrt{Q^2+U^2}/\sigma_{\rm U,V}\approx 1.7$ (where $\sigma_{U,V}=1.7\,\mu$Jy/beam; see Table 1) at 2.8\,GHz and at a mean epoch of $\approx 244$\,d since merger, which implies a 99\% upper-limit on $p$ of $p<3.8$ \citep{Vaillancourt2006}.  The Stokes I peak brightness measured for GW170817 in the co-added image is of $(51.9\pm3.3)\,\mu$Jy/beam \citep[fully consistent with the turnover trend identified by][]{Dobie2018}. Thus, our most stringent upper-limit on the linear polarization fraction of GW170817 is of $\Pi=\sqrt{Q^2+U^2}/I\lesssim 12\%$ at $\approx 244$\,d since merger (Fig. \ref{Fig:pol}). 

As discussed in Section \ref{sec:intro}, a successful structured jet (scenario (i)) and a choked jet - cocoon system (scenario (ii)) can both explain the observed radio light curve of GW170817 \citep[e.g.,][]{Hallinan2017,Kasliwal2017,Gill2018,Lazzati2018,Margutti2018,Mooley2018,Nakar2018a}. 
Thus, polarization observations have been proposed as a way to break this degeneracy and discriminate between scenarios (i) and (ii) \citep{Lazzati2018,Gill2018,Nakar2018b}. The predictions for the linear polarization near the peak of the radio light curve are indeed substantially different in these two cases. For a given magnetic field configuration, the successful jet scenario produces a larger polarization than that expected for a quasi-spherical outflow.
However, for both outflow structures, the predicted polarization fraction also depends strongly on the configuration of the magnetic field (which is usually assumed to be completely tangled in the
plane of the shock). Specifically, the degree of linear polarization is maximum for a magnetic field fully contained within the plane of the shock, and decreases with an increasing magnetic field component in the direction of the shock normal. This effect can be parametrized by the ratio $b=2<B^2_{\rm sn}>/<B^2_{\rm sp}>$, where $b=0$ is for a magnetic field fully contained in the plane of the shock, while for $b>0$ the component of the field along the shock normal also contributes to the emission \citep{Gill2018}.

In Figure \ref{Fig:pol} we show the predictions by \citet{Gill2018}  for the linear polarization fraction of the radio continuum from a successful structured jet with a power-law distribution of energy and Lorentz factors (PLJ; black lines). Similar predictions for the case of a radially stratified quasi-spherical (QS) ejecta are also shown (blue dotted line). These predictions strongly depend on the value of $b$, but for the QS ejecta case we only show $b=0$ as larger values of $b$ would imply even smaller degrees of linear polarization. 

As evident from Figure \ref{Fig:pol}, our VLA linear polarization fraction upper-limit (downward pointing triangle) excludes a structured jet model with $b=0$, and requires $b> 0.5$. We note that analogous magnetic field configurations are inferred comparing optical data of long duration GRBs with various predictions for the expected optical polarization fraction. Analytic models of long GRBs \citep{Ghisellini1999,Sari1999} predict polarization values peaking at $\sim 10-15\%$ (assuming $b=0$) for the most probable observing configuration (with the observer at the edge of the jet). Optical observations, on the other hand, have provided polarization values of long GRBs optical afterglows clustered around a few $\%$ \citep{Covino2016}, even for bursts with multiple observations spanning the entire afterglow evolution \citep[e.g.,][]{Greiner2003,Wiersema2012}. Finally, we note that our upper-limit on the radio linear polarization fraction of GW170817 cannot constrain the QS ejecta scenario, even in the most optimistic case of a magnetic field fully confined in the plane of the shock ($b=0$). 

\section{Summary and Conclusion}
\label{sec:conclusion}
We have presented the first observational constraint on the $\approx 3$\,GHz linear polarization fraction of GW170817 radio continuum. Thanks to the excellent point source sensitivity of the VLA, we are able to unambiguously rule out the most optimistic predictions for the linearly polarized radio flux expected within a successful structured jet scenario. We have also shown that, under the hypothesis that a successful structured jet did indeed form in GW170817, the magnetic field behind the shock cannot be fully contained within the plane of the shock. Instead, a significant component of the magnetic field in the direction perpendicular to the shock plane is required to reconcile theoretical predictions with our observational upper-limit. 

Even though GW170817 is a relatively nearby event ($d_L\approx 40$\,Mpc), its faint radio emission ($\sim10^{26}$\,erg\,s$^{-1}$\,Hz$^{-1}$ peak spectral luminosity density at $\approx 3$\,GHz) does not allow us to constrain the degree of linear polarization of the radio continuum down to a level that could probe the quasi-spherical ejecta formed in a choked jet scenario. More stringent constraints on this scenario may be achieved via direct VLBI imaging of the merger ejecta.

With the advanced LIGO and Virgo detectors now scheduled to start their third observing run \citep[O3;][]{Abbott2016}, we expect to have soon more opportunities for probing the possible variety of radio afterglows from NS-NS (or black hole - NS) mergers. The luminosity of the radio counterparts of these events will depend on several factors, including the ISM density, the total energy in the fastest ejecta, and the viewing geometry. If a compact binary merger with a radio afterglow twice as bright as GW170817 were to be discovered, the VLA could probe linear polarization fractions below $\approx 10\%$. At this level, the absence of linear polarization is likely to challenge strongly the successful structured jet hypothesis (assuming an outflow geometry and viewing angle similar to that of GW170817). 

Finally, looking further into the future, when advanced LIGO and Virgo will be reaching their nominal sensitivities and discovering NS-NS mergers up to $\approx 120-190$\,Mpc \citep{Abbott2016}, detecting radio counterparts as faint as GW170817 will require radio arrays $\approx 10\times$ more sensitive than the VLA. In this respect, the next generation Very Large Array \citep[ngVLA;][]{Murphy2017} may offer us a unique opportunity to probe the different ejecta structures via radio polarimetry. 

\acknowledgements
The National Radio Astronomy Observatory is a facility of the National Science Foundation operated under cooperative agreement by Associated Universities, Inc. We thank the VLA director's office and schedsoc for approving and promptly executing the observations presented in this study. A.C. and D.C. acknowledge support from the National Science Foundation (NSF) CAREER award \#1455090. A.C. thanks Drew Medlin and and Jose Salcido for helpful discussions on the VLA automated calibration pipeline, and on polarization calibration in CASA. A.C. is grateful to Roger Blandford for suggesting to carry out radio polarization observations of GW170817 in the radio. D.K. was additionally supported by NSF grant AST-1412421. D.L. acknowledges support form the NASA ATP grant  NNX17AK42G. K.P.M. is currently a Jansky Fellow of the National Radio Astronomy Observatory.


\begin{thebibliography}{}
\expandafter\ifx\csname natexlab\endcsname\relax\def\natexlab#1{#1}\fi

\end{thebibliography}


\begin{thebibliography}{}
\expandafter\ifx\csname natexlab\endcsname\relax\def\natexlab#1{#1}\fi

\bibitem[{{Abbott} {et~al.}(2016){Abbott}, {Abbott}, {Abbott}, {Abernathy},
  {Acernese}, {Ackley}, {Adams}, {Adams}, {Addesso}, {Adhikari}, \&
  et~al.}]{Abbott2016}
{Abbott}, B.~P., {Abbott}, R., {Abbott}, T.~D., {et~al.} 2016, Living Reviews
  in Relativity, 19, 1

\bibitem[Abbott {et~al.}(2017{\natexlab{a}})]{Abbott2017a}
Abbott, B.~P., Abbott, R., Abbott, T.~D., {et~al.} 2017a, Phys. Rev. Lett., 119,
  161101

\bibitem[{Abbott} {et~al.}(2017{\natexlab{b}})]{Abbott2017b}
{Abbott}, B.~P., {Abbott}, R., {Abbott}, T.~D., {et~al.} 2017b, \apjl, 848, L12

\bibitem[{Abbott} {et~al.}(2017{\natexlab{c}})]{Abbott2017c}{Abbott}, B.~P., {Abbott}, R., {Abbott}, T.~D., {et~al.} 2017c, \apjl, 848, L13


\bibitem[{{Alexander} {et~al.}(2017){Alexander}, {Berger}, {Fong}, {Williams},
  {Guidorzi}, {Margutti}, {Metzger}, {Annis}, {Blanchard}, {Brout}, {Brown},
  {Chen}, {Chornock}, {Cowperthwaite}, {Drout}, {Eftekhari}, {Frieman}, {Holz},
  {Nicholl}, {Rest}, {Sako}, {Soares-Santos}, \& {Villar}}]{Alexander2017}
{Alexander}, K.~D., {Berger}, E., {Fong}, W., {et~al.} 2017, \apjl, 848, L21

\bibitem[{{Alexander} {et~al.}(2018){Alexander}, {Margutti}, {Blanchard},
  {Fong}, {Berger}, {Hajela}, {Eftekhari}, {Chornock}, {Cowperthwaite},
  {Giannios}, {Guidorzi}, {Kathirgamaraju}, {MacFadyen}, {Metzger}, {Nicholl},
  {Sironi}, {Villar}, {Williams}, {Xie}, \& {Zrake}}]{Alexander2018}
{Alexander}, K.~D., {Margutti}, R., {Blanchard}, P.~K., {et~al.} 2018, ArXiv
  e-prints, arXiv:1805.02870

\bibitem[{{Arcavi} {et~al.}(2017){Arcavi}, {Hosseinzadeh}, {Howell}, {McCully},
  {Poznanski}, {Kasen}, {Barnes}, {Zaltzman}, {Vasylyev}, {Maoz}, \&
  {Valenti}}]{Arcavi2017}
{Arcavi}, I., {Hosseinzadeh}, G., {Howell}, D.~A., {et~al.} 2017, \nat, 551, 64


\bibitem[{{Chornock} {et~al.}(2017){Chornock}, {Berger}, {Kasen},
  {Cowperthwaite}, {Nicholl}, {Villar}, {Alexander}, {Blanchard}, {Eftekhari},
  {Fong}, {Margutti}, {Williams}, {Annis}, {Brout}, {Brown}, {Chen}, {Drout},
  {Farr}, {Foley}, {Frieman}, {Fryer}, {Herner}, {Holz}, {Kessler}, {Matheson},
  {Metzger}, {Quataert}, {Rest}, {Sako}, {Scolnic}, {Smith}, \&
  {Soares-Santos}}]{Chornock2017}
{Chornock}, R., {Berger}, E., {Kasen}, D., {et~al.} 2017, \apjl, 848, L19

\bibitem[{Covino} \& {Gotz} (2016)]{Covino2016}{Covino}, S. and {Gotz}, D. A\&AT, 29, 205

\bibitem[{{Coulter} {et~al.}(2017){Coulter}, {Foley}, {Kilpatrick}, {Drout},
  {Piro}, {Shappee}, {Siebert}, {Simon}, {Ulloa}, {Kasen}, {Madore},
  {Murguia-Berthier}, {Pan}, {Prochaska}, {Ramirez-Ruiz}, {Rest}, \&
  {Rojas-Bravo}}]{Coulter2017}
{Coulter}, D.~A., {Foley}, R.~J., {Kilpatrick}, C.~D., {et~al.} 2017, Science,
  358, 1556

\bibitem[{{Cowperthwaite} {et~al.}(2017){Cowperthwaite}, {Berger}, {Villar},
  {Metzger}, {Nicholl}, {Chornock}, {Blanchard}, {Fong}, {Margutti},
  {Soares-Santos}, {Alexander}, {Allam}, {Annis}, {Brout}, {Brown}, {Butler},
  {Chen}, {Diehl}, {Doctor}, {Drout}, {Eftekhari}, {Farr}, {Finley}, {Foley},
  {Frieman}, {Fryer}, {Garc{\'{\i}}a-Bellido}, {Gill}, {Guillochon}, {Herner},
  {Holz}, {Kasen}, {Kessler}, {Marriner}, {Matheson}, {Neilsen}, {Quataert},
  {Palmese}, {Rest}, {Sako}, {Scolnic}, {Smith}, {Tucker}, {Williams},
  {Balbinot}, {Carlin}, {Cook}, {Durret}, {Li}, {Lopes}, {Louren{\c c}o},
  {Marshall}, {Medina}, {Muir}, {Mu{\~n}oz}, {Sauseda}, {Schlegel}, {Secco},
  {Vivas}, {Wester}, {Zenteno}, {Zhang}, {Abbott}, {Banerji}, {Bechtol},
  {Benoit-L{\'e}vy}, {Bertin}, {Buckley-Geer}, {Burke}, {Capozzi}, {Carnero
  Rosell}, {Carrasco Kind}, {Castander}, {Crocce}, {Cunha}, {D'Andrea}, {da
  Costa}, {Davis}, {DePoy}, {Desai}, {Dietrich}, {Drlica-Wagner}, {Eifler},
  {Evrard}, {Fernandez}, {Flaugher}, {Fosalba}, {Gaztanaga}, {Gerdes},
  {Giannantonio}, {Goldstein}, {Gruen}, {Gruendl}, {Gutierrez}, {Honscheid},
  {Jain}, {James}, {Jeltema}, {Johnson}, {Johnson}, {Kent}, {Krause}, {Kron},
  {Kuehn}, {Nuropatkin}, {Lahav}, {Lima}, {Lin}, {Maia}, {March}, {Martini},
  {McMahon}, {Menanteau}, {Miller}, {Miquel}, {Mohr}, {Neilsen}, {Nichol},
  {Ogando}, {Plazas}, {Roe}, {Romer}, {Roodman}, {Rykoff}, {Sanchez},
  {Scarpine}, {Schindler}, {Schubnell}, {Sevilla-Noarbe}, {Smith}, {Smith},
  {Sobreira}, {Suchyta}, {Swanson}, {Tarle}, {Thomas}, {Thomas}, {Troxel},
  {Vikram}, {Walker}, {Wechsler}, {Weller}, {Yanny}, \&
  {Zuntz}}]{Cowperthwaite2017}
{Cowperthwaite}, P.~S., {Berger}, E., {Villar}, V.~A., {et~al.} 2017, \apjl,
  848, L17

\bibitem[{{Dobie} {et~al.}(2018){Dobie}, {Kaplan}, {Murphy}, {Lenc}, {Mooley},
  {Lynch}, {Corsi}, {Frail}, {Kasliwal}, \& {Hallinan}}]{Dobie2018}
{Dobie}, D., {Kaplan}, D.~L., {Murphy}, T., {et~al.} 2018, \apjl, 858, 15

\bibitem[{{Drout} {et~al.}(2017){Drout}, {Piro}, {Shappee}, {Kilpatrick},
  {Simon}, {Contreras}, {Coulter}, {Foley}, {Siebert}, {Morrell}, {Boutsia},
  {Di Mille}, {Holoien}, {Kasen}, {Kollmeier}, {Madore}, {Monson},
  {Murguia-Berthier}, {Pan}, {Prochaska}, {Ramirez-Ruiz}, {Rest}, {Adams},
  {Alatalo}, {Ba{\~n}ados}, {Baughman}, {Beers}, {Bernstein}, {Bitsakis},
  {Campillay}, {Hansen}, {Higgs}, {Ji}, {Maravelias}, {Marshall}, {Bidin},
  {Prieto}, {Rasmussen}, {Rojas-Bravo}, {Strom}, {Ulloa},
  {Vargas-Gonz{\'a}lez}, {Wan}, \& {Whitten}}]{Drout2017}
{Drout}, M.~R., {Piro}, A.~L., {Shappee}, B.~J., {et~al.} 2017, Science, 358,
  1570


\bibitem[{{Evans} {et~al.}(2017){Evans}, {Cenko}, {Kennea}, {Emery}, {Kuin},
  {Korobkin}, {Wollaeger}, {Fryer}, {Madsen}, {Harrison}, {Xu}, {Nakar},
  {Hotokezaka}, {Lien}, {Campana}, {Oates}, {Troja}, {Breeveld}, {Marshall},
  {Barthelmy}, {Beardmore}, {Burrows}, {Cusumano}, {D'A{\`i}}, {D'Avanzo},
  {D'Elia}, {de Pasquale}, {Even}, {Fontes}, {Forster}, {Garcia}, {Giommi},
  {Grefenstette}, {Gronwall}, {Hartmann}, {Heida}, {Hungerford}, {Kasliwal},
  {Krimm}, {Levan}, {Malesani}, {Melandri}, {Miyasaka}, {Nousek}, {O'Brien},
  {Osborne}, {Pagani}, {Page}, {Palmer}, {Perri}, {Pike}, {Racusin}, {Rosswog},
  {Siegel}, {Sakamoto}, {Sbarufatti}, {Tagliaferri}, {Tanvir}, \&
  {Tohuvavohu}}]{Evans2017}
{Evans}, P.~A., {Cenko}, S.~B., {Kennea}, J.~A., {et~al.} 2017, Science, 358,
  1565

\bibitem[{{Fong} {et~al.}(2017){Fong}, {Berger}, {Blanchard}, {Margutti},
  {Cowperthwaite}, {Chornock}, {Alexander}, {Metzger}, {Villar}, {Nicholl},
  {Eftekhari}, {Williams}, {Annis}, {Brout}, {Brown}, {Chen}, {Doctor},
  {Diehl}, {Holz}, {Rest}, {Sako}, \& {Soares-Santos}}]{Fong2017}
{Fong}, W., {Berger}, E., {Blanchard}, P.~K., {et~al.} 2017, \apjl, 848, L23

\bibitem[{Ghisellini} \& {Lazzati} (1999)]{Ghisellini1999}{Ghisellini}, G. and {Lazzati}, D. 1999, \mnras, 309, L7

\bibitem[{{Gill} \& {Granot}(2018)}]{Gill2018}
{Gill}, R., \& {Granot}, J. 2018, ArXiv e-prints, arXiv:1803.05892

\bibitem[{{Goldstein} {et~al.}(2017){Goldstein}, {Veres}, {Burns}, {Briggs},
  {Hamburg}, {Kocevski}, {Wilson-Hodge}, {Preece}, {Poolakkil}, {Roberts},
  {Hui}, {Connaughton}, {Racusin}, {von Kienlin}, {Dal Canton}, {Christensen},
  {Littenberg}, {Siellez}, {Blackburn}, {Broida}, {Bissaldi}, {Cleveland},
  {Gibby}, {Giles}, {Kippen}, {McBreen}, {McEnery}, {Meegan}, {Paciesas}, \&
  {Stanbro}}]{Goldstein2017}
{Goldstein}, A., {Veres}, P., {Burns}, E., {et~al.} 2017, \apjl, 848, L14

\bibitem[{{Gottlieb} {et~al.}(2018){Gottlieb}, {Nakar}, \&
  {Piran}}]{Gottlieb2018}
{Gottlieb}, O., {Nakar}, E., \& {Piran}, T. 2018, \mnras, 473, 576

\bibitem[{Greiner} et ~al. (2003)]{Greiner2003} {Greiner}, J. and {Klose}, S. and {Reinsch}, K., {et~al.} 2003, \nat, 426, 157

\bibitem[{{Haggard} {et~al.}(2017){Haggard}, {Nynka}, {Ruan}, {Kalogera},
  {Cenko}, {Evans}, \& {Kennea}}]{Haggard2017}
{Haggard}, D., {Nynka}, M., {Ruan}, J.~J., {et~al.} 2017, \apjl, 848, L25

\bibitem[{{Hallinan}, {Corsi} {et~al.}(2017){Hallinan}, {Corsi}, {Mooley}, {Hotokezaka},
  {Nakar}, {Kasliwal}, {Kaplan}, {Frail}, {Myers}, {Murphy}, {De}, {Dobie},
  {Allison}, {Bannister}, {Bhalerao}, {Chandra}, {Clarke}, {Giacintucci}, {Ho},
  {Horesh}, {Kassim}, {Kulkarni}, {Lenc}, {Lockman}, {Lynch}, {Nichols},
  {Nissanke}, {Palliyaguru}, {Peters}, {Piran}, {Rana}, {Sadler}, \&
  {Singer}}]{Hallinan2017}
{Hallinan}, G., {Corsi}, A., {Mooley}, K.~P., {et~al.} 2017, Science, 358, 1579

\bibitem[{{H{\"o}gbom}(1974)}]{Hogbom1974}
{H{\"o}gbom}, J.~A. 1974, \aaps, 15, 417

\bibitem[Hotokezaka et al. (2018)]{Hotokezaka2018} Hotokezaka, K., Kiuchi, K.,  Shibata, M.,  Nakar, E., Piran, T. 2018, eprint arXiv:1803.00599

\bibitem[{{Kasen} {et~al.}(2017){Kasen}, {Metzger}, {Barnes}, {Quataert}, \&
  {Ramirez-Ruiz}}]{Kasen2017}
{Kasen}, D., {Metzger}, B., {Barnes}, J., {Quataert}, E., \& {Ramirez-Ruiz}, E.
  2017, \nat, 551, 80

\bibitem[{{Kasliwal} {et~al.}(2017){Kasliwal}, {Nakar}, {Singer}, {Kaplan},
  {Cook}, {Van Sistine}, {Lau}, {Fremling}, {Gottlieb}, {Jencson}, {Adams},
  {Feindt}, {Hotokezaka}, {Ghosh}, {Perley}, {Yu}, {Piran}, {Allison},
  {Anupama}, {Balasubramanian}, {Bannister}, {Bally}, {Barnes}, {Barway},
  {Bellm}, {Bhalerao}, {Bhattacharya}, {Blagorodnova}, {Bloom}, {Brady},
  {Cannella}, {Chatterjee}, {Cenko}, {Cobb}, {Copperwheat}, {Corsi}, {De},
  {Dobie}, {Emery}, {Evans}, {Fox}, {Frail}, {Frohmaier}, {Goobar}, {Hallinan},
  {Harrison}, {Helou}, {Hinderer}, {Ho}, {Horesh}, {Ip}, {Itoh}, {Kasen},
  {Kim}, {Kuin}, {Kupfer}, {Lynch}, {Madsen}, {Mazzali}, {Miller}, {Mooley},
  {Murphy}, {Ngeow}, {Nichols}, {Nissanke}, {Nugent}, {Ofek}, {Qi}, {Quimby},
  {Rosswog}, {Rusu}, {Sadler}, {Schmidt}, {Sollerman}, {Steele}, {Williamson},
  {Xu}, {Yan}, {Yatsu}, {Zhang}, \& {Zhao}}]{Kasliwal2017}
{Kasliwal}, M.~M., {Nakar}, E., {Singer}, L.~P., {et~al.} 2017, Science, 358,
  1559

\bibitem[{{Kilpatrick} {et~al.}(2017){Kilpatrick}, {Foley}, {Kasen},
  {Murguia-Berthier}, {Ramirez-Ruiz}, {Coulter}, {Drout}, {Piro}, {Shappee},
  {Boutsia}, {Contreras}, {Di Mille}, {Madore}, {Morrell}, {Pan}, {Prochaska},
  {Rest}, {Rojas-Bravo}, {Siebert}, {Simon}, \& {Ulloa}}]{Kilpatrick2017}
{Kilpatrick}, C.~D., {Foley}, R.~J., {Kasen}, D., {et~al.} 2017, Science, 358,
  1583

\bibitem[{{Lazzati} {et~al.}(2017{\natexlab{a}}){Lazzati}, {Deich}, {Morsony},
  \& {Workman}}]{Lazzati2017a}
{Lazzati}, D., {Deich}, A., {Morsony}, B.~J., \& {Workman}, J.~C.
  2017{\natexlab{a}}, \mnras, 471, 1652

\bibitem[{{Lazzati} {et~al.}(2017{\natexlab{b}}){Lazzati},
  {L{\'o}pez-C{\'a}mara}, {Cantiello}, {Morsony}, {Perna}, \&
  {Workman}}]{Lazzati2017b}
{Lazzati}, D., {L{\'o}pez-C{\'a}mara}, D., {Cantiello}, M., {et~al.}
  2017{\natexlab{b}}, \apjl, 848, L6

\bibitem[{{Lazzati} {et~al.}(2017{\natexlab{c}}){Lazzati}, {Perna}, {Morsony},
  {L{\'o}pez-C{\'a}mara}, {Cantiello}, {Ciolfi}, {giacomazzo}, \&
  {Workman}}]{Lazzati2018}
{Lazzati}, D., {Perna}, R., {Morsony}, B.~J., {et~al.} 2017{\natexlab{c}},
  ArXiv e-prints, arXiv:1712.03237

\bibitem[{{Margutti} {et~al.}(2017){Margutti}, {Berger}, {Fong}, {Guidorzi},
  {Alexander}, {Metzger}, {Blanchard}, {Cowperthwaite}, {Chornock},
  {Eftekhari}, {Nicholl}, {Villar}, {Williams}, {Annis}, {Brown}, {Chen},
  {Doctor}, {Frieman}, {Holz}, {Sako}, \& {Soares-Santos}}]{Margutti2017}
{Margutti}, R., {Berger}, E., {Fong}, W., {et~al.} 2017, \apjl, 848, L20

\bibitem[{{Margutti} {et~al.}(2018){Margutti}, {Alexander}, {Xie}, {Sironi},
  {Metzger}, {Kathirgamaraju}, {Fong}, {Blanchard}, {Berger}, {MacFadyen},
  {Giannios}, {Guidorzi}, {Hajela}, {Chornock}, {Cowperthwaite}, {Eftekhari},
  {Nicholl}, {Villar}, {Williams}, \& {Zrake}}]{Margutti2018}
{Margutti}, R., {Alexander}, K.~D., {Xie}, X., {et~al.} 2018, \apjl, 856, L18


\bibitem[{{Metzger}(2017)}]{Metzger2017}
{Metzger}, B.~D. 2017, ArXiv e-prints, arXiv:1710.05931

\bibitem[{{Mooley} {et~al.}(2018){Mooley}, {Nakar}, {Hotokezaka}, {Hallinan},
  {Corsi}, {Frail}, {Horesh}, {Murphy}, {Lenc}, {Kaplan}, {de}, {Dobie},
  {Chandra}, {Deller}, {Gottlieb}, {Kasliwal}, {Kulkarni}, {Myers}, {Nissanke},
  {Piran}, {Lynch}, {Bhalerao}, {Bourke}, {Bannister}, \&
  {Singer}}]{Mooley2018}
{Mooley}, K.~P., {Nakar}, E., {Hotokezaka}, K., {et~al.} 2018, \nat, 554, 207

\bibitem[{Murphy} (2017)]{Murphy2017}
{Murphy}, E. 2017, ArXiv e-prints, arXiv:1711.09921

\bibitem[{{Nakar} {et~al.}(2018){Nakar}, {Gottlieb}, {Piran}, {Kasliwal}, \&
  {Hallinan}}]{Nakar2018b}
{Nakar}, E., {Gottlieb}, O., {Piran}, T., {Kasliwal}, M.~M., \& {Hallinan}, G.
  2018, ArXiv e-prints, arXiv:1803.07595

\bibitem[{{Nakar} \& {Piran}(2017)}]{Nakar2017}
{Nakar}, E., \& {Piran}, T. 2017, \apj, 834, 28

\bibitem[{{Nakar} \& {Piran}(2018)}]{Nakar2018a}
---. 2018, \mnras, arXiv:1801.09712


\bibitem[{{Perley} \& {Butler}(2013)}]{Perley2013}
{Perley}, R.~A., \& {Butler}, B.~J. 2013, \apjs, 206, 16

\bibitem[{{Pian} {et~al.}(2017){Pian}, {D'Avanzo}, {Benetti}, {Branchesi},
  {Brocato}, {Campana}, {Cappellaro}, {Covino}, {D'Elia}, {Fynbo}, {Getman},
  {Ghirlanda}, {Ghisellini}, {Grado}, {Greco}, {Hjorth}, {Kouveliotou},
  {Levan}, {Limatola}, {Malesani}, {Mazzali}, {Melandri}, {M{\o}ller},
  {Nicastro}, {Palazzi}, {Piranomonte}, {Rossi}, {Salafia}, {Selsing},
  {Stratta}, {Tanaka}, {Tanvir}, {Tomasella}, {Watson}, {Yang}, {Amati},
  {Antonelli}, {Ascenzi}, {Bernardini}, {Bo{\"e}r}, {Bufano}, {Bulgarelli},
  {Capaccioli}, {Casella}, {Castro-Tirado}, {Chassande-Mottin}, {Ciolfi},
  {Copperwheat}, {Dadina}, {De Cesare}, {di Paola}, {Fan}, {Gendre},
  {Giuffrida}, {Giunta}, {Hunt}, {Israel}, {Jin}, {Kasliwal}, {Klose}, {Lisi},
  {Longo}, {Maiorano}, {Mapelli}, {Masetti}, {Nava}, {Patricelli}, {Perley},
  {Pescalli}, {Piran}, {Possenti}, {Pulone}, {Razzano}, {Salvaterra},
  {Schipani}, {Spera}, {Stamerra}, {Stella}, {Tagliaferri}, {Testa}, {Troja},
  {Turatto}, {Vergani}, \& {Vergani}}]{Pian2017}
{Pian}, E., {D'Avanzo}, P., {Benetti}, S., {et~al.} 2017, \nat, 551, 67

\bibitem[{{Rossi} {et~al.}(2004){Rossi}, {Lazzati}, {Salmonson}, \&
  {Ghisellini}}]{Rossi2004}
{Rossi}, E.~M., {Lazzati}, D., {Salmonson}, J.~D., \& {Ghisellini}, G. 2004,
  \mnras, 354, 86

\bibitem[{Sari} (1999)]{Sari1999}{Sari}, R. 1999, \apjl, 524, L43

\bibitem[{{Savchenko} {et~al.}(2017){Savchenko}, {Ferrigno}, {Kuulkers},
  {Bazzano}, {Bozzo}, {Brandt}, {Chenevez}, {Courvoisier}, {Diehl}, {Domingo},
  {Hanlon}, {Jourdain}, {von Kienlin}, {Laurent}, {Lebrun}, {Lutovinov},
  {Martin-Carrillo}, {Mereghetti}, {Natalucci}, {Rodi}, {Roques}, {Sunyaev}, \&
  {Ubertini}}]{Savchenko2017}
{Savchenko}, V., {Ferrigno}, C., {Kuulkers}, E., {et~al.} 2017, \apjl, 848, L15

\bibitem[{{Shappee} {et~al.}(2017){Shappee}, {Simon}, {Drout}, {Piro},
  {Morrell}, {Prieto}, {Kasen}, {Holoien}, {Kollmeier}, {Kelson}, {Coulter},
  {Foley}, {Kilpatrick}, {Siebert}, {Madore}, {Murguia-Berthier}, {Pan},
  {Prochaska}, {Ramirez-Ruiz}, {Rest}, {Adams}, {Alatalo}, {Ba{\~n}ados},
  {Baughman}, {Bernstein}, {Bitsakis}, {Boutsia}, {Bravo}, {Di Mille}, {Higgs},
  {Ji}, {Maravelias}, {Marshall}, {Placco}, {Prieto}, \& {Wan}}]{Shappee2017}
{Shappee}, B.~J., {Simon}, J.~D., {Drout}, M.~R., {et~al.} 2017, Science, 358,
  1574
  

\bibitem[{{Smartt} {et~al.}(2017){Smartt}, {Chen}, {Jerkstrand}, {Coughlin},
  {Kankare}, {Sim}, {Fraser}, {Inserra}, {Maguire}, {Chambers}, {Huber},
  {Kr{\"u}hler}, {Leloudas}, {Magee}, {Shingles}, {Smith}, {Young}, {Tonry},
  {Kotak}, {Gal-Yam}, {Lyman}, {Homan}, {Agliozzo}, {Anderson}, {Angus},
  {Ashall}, {Barbarino}, {Bauer}, {Berton}, {Botticella}, {Bulla}, {Bulger},
  {Cannizzaro}, {Cano}, {Cartier}, {Cikota}, {Clark}, {De Cia}, {Della Valle},
  {Denneau}, {Dennefeld}, {Dessart}, {Dimitriadis}, {Elias-Rosa}, {Firth},
  {Flewelling}, {Fl{\"o}rs}, {Franckowiak}, {Frohmaier}, {Galbany},
  {Gonz{\'a}lez-Gait{\'a}n}, {Greiner}, {Gromadzki}, {Guelbenzu},
  {Guti{\'e}rrez}, {Hamanowicz}, {Hanlon}, {Harmanen}, {Heintz}, {Heinze},
  {Hernandez}, {Hodgkin}, {Hook}, {Izzo}, {James}, {Jonker}, {Kerzendorf},
  {Klose}, {Kostrzewa-Rutkowska}, {Kowalski}, {Kromer}, {Kuncarayakti},
  {Lawrence}, {Lowe}, {Magnier}, {Manulis}, {Martin-Carrillo}, {Mattila},
  {McBrien}, {M{\"u}ller}, {Nordin}, {O'Neill}, {Onori}, {Palmerio},
  {Pastorello}, {Patat}, {Pignata}, {Podsiadlowski}, {Pumo}, {Prentice}, {Rau},
  {Razza}, {Rest}, {Reynolds}, {Roy}, {Ruiter}, {Rybicki}, {Salmon}, {Schady},
  {Schultz}, {Schweyer}, {Seitenzahl}, {Smith}, {Sollerman}, {Stalder},
  {Stubbs}, {Sullivan}, {Szegedi}, {Taddia}, {Taubenberger}, {Terreran}, {van
  Soelen}, {Vos}, {Wainscoat}, {Walton}, {Waters}, {Weiland}, {Willman},
  {Wiseman}, {Wright}, {Wyrzykowski}, \& {Yaron}}]{Smartt2017}
{Smartt}, S.~J., {Chen}, T.-W., {Jerkstrand}, A., {et~al.} 2017, \nat, 551, 75

\bibitem[{{Tanvir} {et~al.}(2017){Tanvir}, {Levan},
  {Gonz{\'a}lez-Fern{\'a}ndez}, {Korobkin}, {Mandel}, {Rosswog}, {Hjorth},
  {D'Avanzo}, {Fruchter}, {Fryer}, {Kangas}, {Milvang-Jensen}, {Rosetti},
  {Steeghs}, {Wollaeger}, {Cano}, {Copperwheat}, {Covino}, {D'Elia}, {de Ugarte
  Postigo}, {Evans}, {Even}, {Fairhurst}, {Figuera Jaimes}, {Fontes}, {Fujii},
  {Fynbo}, {Gompertz}, {Greiner}, {Hodosan}, {Irwin}, {Jakobsson},
  {J{\o}rgensen}, {Kann}, {Lyman}, {Malesani}, {McMahon}, {Melandri},
  {O'Brien}, {Osborne}, {Palazzi}, {Perley}, {Pian}, {Piranomonte}, {Rabus},
  {Rol}, {Rowlinson}, {Schulze}, {Sutton}, {Th{\"o}ne}, {Ulaczyk}, {Watson},
  {Wiersema}, \& {Wijers}}]{Tanvir2017}
{Tanvir}, N.~R., {Levan}, A.~J., {Gonz{\'a}lez-Fern{\'a}ndez}, C., {et~al.}
  2017, \apjl, 848, L27

\bibitem[{{Troja} {et~al.}(2017){Troja}, {Piro}, {van Eerten}, {Wollaeger},
  {Im}, {Fox}, {Butler}, {Cenko}, {Sakamoto}, {Fryer}, {Ricci}, {Lien}, {Ryan},
  {Korobkin}, {Lee}, {Burgess}, {Lee}, {Watson}, {Choi}, {Covino}, {D'Avanzo},
  {Fontes}, {Gonz{\'a}lez}, {Khandrika}, {Kim}, {Kim}, {Lee}, {Lee}, {Kutyrev},
  {Lim}, {S{\'a}nchez-Ram{\'{\i}}rez}, {Veilleux}, {Wieringa}, \&
  {Yoon}}]{Troja2017}
{Troja}, E., {Piro}, L., {van Eerten}, H., {et~al.} 2017, \nat, 551, 71

\bibitem[Vaillancourt (2006)]{Vaillancourt2006} Vaillancourt, J.~E. 2006, \pasp, 118, 1340

\bibitem[{{Valenti} {et~al.}(2017){Valenti}, {David}, {Sand}, {Yang},
  {Cappellaro}, {Tartaglia}, {Corsi}, {Jha}, {Reichart}, {Haislip}, \&
  {Kouprianov}}]{Valenti2017}
{Valenti}, S., {David}, {Sand}, J., {et~al.} 2017, \apjl, 848, L24

\bibitem[Wiersema {et~al.} (2012)]{Wiersema2012} {Wiersema}, K. and {Curran}, P.~A. and {Kr{\"u}hler}, T., {et~al.} 2012, \mnras, 426, 2

\end{thebibliography}

\end{document}